\documentclass[useAMS,usenatbib]{mn2e}
\title[A generic relation between baryonic and radiative energy densities of stars]{A generic relation between baryonic and radiative energy densities of stars}
\author[A. Mitra ]{A. Mitra$^1$\thanks{amitra@apsara.barc.ernet.in}\\
$^1$ Theoretical Astrophysics Section, Bhabha Atomic Research Centre, Mumbai-400085 India\\}
\begin{document}

\date{Accepted 2006 January 5. Received 2006 January 5; in original form 2005 November 16}

\pagerange{\pageref{firstpage}--\pageref{lastpage}} \pubyear{}

\maketitle

\label{firstpage}

\begin{abstract}
By using elementary astrophysical concepts, we show that for any self-luminous astrophysical object, the ratio of radiation energy density inside the body ($\rho_r$) and the baryonic energy density ($\rho_0$) may be crudely approximated,
in the Newtonian limit, as $\rho_r/\rho_0 \propto GM/Rc^2$, where $G$ is constant of gravitation, 
$c$ is the speed of light, $M$ is gravitational mass, and $R$ is the radius of the body.  The key idea is
that radiation quanta must move out in a  diffusive manner rather than  stream freely inside the body of the star. When one  would move to the extreme General Realtivistic case
i.e., if the surface gravitational redshift, $z \gg 1$, it is found that,
$\rho_r/\rho_0 \propto (1+z)$. Earlier treatments of gravitational collapse, in contrast, generally assumed $\rho_r/\rho_0 \ll 1$. Thus, actually, during continued general relativistic gravitational collapse 
to the Black Hole state ($z \to \infty$), the collapsing matter may essentially 
become an extremely hot fireball with $\rho_r/\rho_0 \gg 1$ {\it a la} the very early Universe even though
the observed luminosity of the body as seen by a faraway observer, $L^\infty \propto (1+z)^{-1}\to 0$ as
$z \to \infty$, and the collapse might appear as ``adiabatic''. 

\end{abstract}


\section{Introduction}
Any object at a finite temperature has a certain radiation energy density. In
particular, we consider here self-luminous astrophysical bodies where the source of
this radiation energy is the intrinisic luminosity $L$ of the body rather
than any externally imposed radiation or accretion.    We show that, since the radiation necessarily diffuses out of the body, the interior radiation density is exceedingly higher than what is found outside the star.  First we consider this problem in the purely Newtonian limit
and find that $\rho_r$ is direcly proportional to the Newtonian compactness $GM/Rc^2$. Since Newtonian compactness 
 is $\ll 1$ for most of the astrophysical objects, naturally, $\rho_r
\ll \rho_0 $, where $\rho_0$
is its baryonic density. However when we move to the extreme General Relativistic case, which
would occur during continued gravitational  collapse to the Black Hole (BH) stage, one may find $\rho_r \gg \rho_0$.

\section{ Newtonian Stars}
If the radiation fluid is moving with a bulk speed $v_{eff}$ with respect to the rest frame of the star, then associated comoving energy
density is
\begin{equation}
\rho_r = {L\over 4\pi R^2 v_{eff}}
\end{equation}
Outside the body, the radiation free streams, and $v_{eff} =c$ irrespective of whether the star is collapsing or not, so that
\begin{equation}
\rho_r^{ex} = {L\over 4\pi R^2 c}
\end{equation}
However, inside the body, the radiation quanta interact with the stellar material
and can only diffuse out. For instance while the free streaming time within the Sun
is $\sim R_\odot/c \sim 2$s, photons created at the core of the Sun may take millions
of years to leave the Sun \citep{b1}. It is because of this slow bulk propagation of photons
by diffusion, $\rho_r$ is exceedingly higher than what is indicated by Eq.(2).
 On the average, the diffusive time scale is \citep{b1}
\begin{equation}
t_d \sim {R^2 \over \lambda c}
\end{equation}
because in between the collisions, the quanta move with a speed of $c$. Here
the mean free path of the quanta is
\begin{equation}
\lambda = (n \sigma)^{-1}
\end{equation}
where $\sigma$ is the appropriate mean matter-radiation interaction cross-section.
We really do not require here any specific value or formula for this $\sigma$ since we
do not want to build any  model. Thus, the $\sigma$ used here may be considered
as a general one. In practical cases, $\sigma$,
can depend on temperature and hence on $L$ itself. But since we would be working
with a {\em ratio of luminosities} rathe than with $L$ itself (see Eq.[9] below),
the present treatment would be approximately valid even in the face of likely complex
behaviour of $\sigma$.
The bulk propagation speed of the radiation inside the star and in the rest frame of the stellar fluid is
\begin{equation}
v_{eff} = {R\over t_d}
\end{equation}
Using Eqs.(3) and (4) in (5), we have
\begin{equation}
v_{eff} \sim {c\over R n \sigma}
\end{equation}
It may be noted that the quantity $\tau = R n \sigma$ is the opacity of stellar body. For
all practical astrophysical cases, $\tau \gg 1$ and $v_{eff} \ll c$.   Accordingly,
from Eqs.(1) and (6), we obtain
\begin{equation}
\rho_r = {L n \sigma \over 4\pi R c}
\end{equation}
Obviously $\rho_r \gg \rho_r^{ex}$.  The Eddington luminosity, i.e., the maximum permissible
luminosity of the star is defined as\citep{b3}
\begin{equation}
L_{ed} = {4 \pi G M c m_p\over \sigma}
\end{equation}
In case {\em one would assume}, $\sigma = \sigma_T$, the Thompson luminosity, the implicit {\em assumption would be} 
that the stellar material is fully ionized hydrogen and the radiation quanta are X-ray photons.
However, we keep here both $L_{ed}$ and $\sigma$ to be unspecified and generic parameters since
we do not require and specific numerical values of $L_{ed}$  or $\sigma$ except for qualitative
illustrations. Let the star be radiating at  a certain fraction $\alpha$ of this generic instantaneous 
critical value:
\begin{equation}
L = \alpha L_{ed}
\end{equation}
Note that while any energy dependence or other complexities in the behaviour of
$\sigma$ would change both $L$ and $L_{ed}$, $\alpha$ itself may be considered
as energy independent.
Then combining Eqs.(7-9), it follows that
\begin{equation}
\rho_r = \alpha {G M m_p n\over R}
\end{equation}
Since the baryonic energy density is $\rho_0 = m_p n c^2$,  the above equation becomes
\begin{equation}
\rho_r = \alpha \rho_0 {G M \over Rc^2}
\end{equation}
Or,
\begin{equation}
{\rho_r\over \rho_0} = \alpha {G M \over Rc^2}
\end{equation}
Thus we see that $\rho_r\over \rho_0$ is directly proportional to the {\em compactness} parameter for a given luminosity. Further this ratio does not depend
on the actual nature of the radiation (i.e., photon or neutrino) or its
interaction property (i.e., $\sigma$) though, the exterior radiation density
 $\rho_r^{ex} \propto \sigma^{-1}$ because $L_{ed} \propto \sigma^{-1}$. 
 For the Sun, $L_\odot \approx 2\times 10^{33}$erg/s, and $\alpha = L_\odot/L_{ed}
\sim 10^{-5}$. Also, the compactness parameter for Sun is $GM/Rc^2 \sim 2\times 10^{-6}$. Therefore, from Eq.(12), we find that Sun has
\begin{equation}
\rho_r/\rho_0 \sim 2\times 10^{-11}
\end{equation}
Since for Sun, $\rho_0 \approx 4\times 10^{21}$erg cm$^{-3}$, Eq.(13) suggests a value of
$\rho_r \sim 8\times 10^{10} $erg cm$^{-3}$. By using the $\rho_r = a T^4$ formula
where $a=7.56 \times 10^{-15}$ cgs is the radiation constant, we obtain a mean
temperature of Sun, $T \sim 2\times 10^6$ K. Since this temperature is obtained
by merely using {\em compactness} of the object, it might be called as ``Compactness
Temperature''. Recalling that the core temperature of Sun is $\sim 2\times 10^7$K, the compactness temperature
obtained here appears to be reasonable.

 During Type II supernova events,
the neutrino luminosity could be $\sim 10^{52}$ erg/s\citep{ b3}, while the corresponding
Eddington value is $\sim 2\times 10^{54}$ erg/s\citep{b3}. Thus in this case of gravitational collapse $\alpha \sim 0.01$. The
 compactness parameter for the just born neutron star could be $\sim 0.1$. Thus in this case,  $\rho_r/\rho_0 \sim 10^{-3}$, which, on a relative solar scale, is a        significantly large number.

It may be mentioned here that self-luminosity is possible under two basic conditions:

(i) Burning of nuclear of other appropriate fuel as is the case for main sequence
stars. In such a case, the astrophysical body can generate self-luminosity even
in the absence of any gravitational contraction.

(ii) Even in the absence of any ignition of nuclear or other fuel, a self-gravitating
system may generate self-luminosity by gravitational contraction by virtue of
negative specific heat associated with self-gravitation \citep{b1}. This is the way primordial
astrophysical clouds remains quasi-stable for millions of years by generating own
pressure, temperature and luminosity by slow gravitational contraction. The originally cold primordial clouds become hot enough to be visible in the optical range and give birth to premain sequence stars. The hot premain sequence stars too
shine by slow gravitational contraction and without ignition of any nuclear fuel.

One might argue here that what if one would consider a {\em perfectly degenerate}
and cold compact object at temperature $T=0$. In such a case, obviously, one would
have $\alpha =0$. However such an idealized object cannot undergo gravitational
contraction because the negative specific heat associated with gravitation demands
that there cannot be any gravitational contraction without the emission of radiation
unless the effective ratio of specific heat of the body is strictly $\gamma =4/3$. It is known however that $\gamma=4/3$ is only an idealization and can happen in three
cases:

$\bullet$ A perfectly degenerate gas having ultimate relativistic degeneracy where
 linear momentum of gas particles, $p \to \infty$ \citep{b1}.

$\bullet$ An incoherent perfect photon gas with no baryonic/rest mass loading so that
conserved energy per unit rest mass $e =\infty$.

$\bullet$ A non-degenerate baryonic/leptonic gas where the lorentz factor of
the atoms is {\em infinite} so that $e= E/m = \infty$. In other words a baryonic
or leptonic fluid must kinemetically behave like a pure photon fluid in order
to have $\gamma = 4/3$.

Eventually all the above cases correspond to particles having either degeneracy related internal Lorentz factor or ordinary thermal Lorentz factor becoming $\infty$.
Since this is an extreme idealization, in all realistic cases $\gamma > 4/3$.

For instance, if the proto neutron star would be considered as perfectly degenerate and cold, it would not be able to collapse. On the other hand, when it is collapsing,
it dictates in own physics and must partially lift the degenearacy and be hot enough
to radiate an appropiate amount of energy. In such a case, one cannot preset the $\alpha =0$
condition, and instead, $\alpha$ would acquire its own appropriate value depending
on various physical aspects in a self-consistent way.
\section{Relativistic Generalization}
The interior of the star may be described by the metric
\begin{equation}
ds^2 = a^2(R,t) dt^2 - b^2(R, t) dR^2 - R^2(d\theta^2 +\sin^2\theta d\phi^2)
\end{equation}
where $\theta$ and $\phi$ are polar and azimuth angles.
In the GR domain while $R$ remains the {\it Luminosity Distance}, in Eqs. (3), one needs to replace $R$ by the proper radius $R_p = b R$, 
i.e., now
\begin{equation}
t_d \sim {b^2 R^2\over \lambda c}
\end{equation}
Corresponding proper time of diffusion is
\begin{equation}
\tau_d \sim a t_d \sim {a b^2 R^2\over \lambda c}
\end{equation}
With these crude modifications, the bulk flow speed of the radiation is
\begin{equation}
v_{eff} \sim {R_p\over \tau_d} \sim {1\over a b} {c\over R n \sigma}
\end{equation}
While for a vacuum Schwarzschild metric, $a b =1$(exactly), in the presence
of mass energy $a b \sim 1$, for instance, for a uniform density 
star, $a b =  1/2$.
 Since we are interested only in crude
qualitative estimates, we ignore such differences in the value of $a b$. 
Hence, 
 the eventual expression for $v_{eff}$ in the GR case, practically, remains unaltered.
However, the expression for {\em local} Eddington luminosity gets modified in GR\citep{b2, b3}:
\begin{equation}
L_{ed} = {4 \pi G M c m_p\over \sigma} (1+z)
\end{equation}
where the surface gravitational redshift of the star is defined as
\begin{equation}
 z= (1- 2 GM/Rc^2)^{-1/2} -1
 \end{equation}
 In fact, in GR, $z$ is the appropriate compactness parameter and it can be easily seen
 that when $GM/Rc^2 \ll 1$, (i.e., in the truly Newtonian regime), one has
 \begin{equation}
 z \approx GM/Rc^2
 \end{equation}

As before we parameterize the actual luminosity of the star by means of Eq.(9).  
Then combining Eqs. (1), (9) and (17), it follows that
\begin{equation}
{\rho_r\over \rho_0} \sim \alpha {G M \over Rc^2} (1+z)
\end{equation}
From Eq.(19), we can write
\begin{equation}
{2 GM\over Rc^2} = 1 - (1+z)^{-2}
\end{equation}
and then rewrite Eq.(21) in a more appealing form
\begin{equation}
{\rho_r\over \rho_0} \sim {\alpha\over 2} [(1+z)  -(1+z)^{-1}]
\end{equation}

In the Newtonian limit of $z \ll 1$, it follows then that
\begin{equation}
{\rho_r\over \rho_0} \sim \alpha z
\end{equation}
And in the extreme relativistic case of $z\gg 1$, we again have
\begin{equation}
{\rho_r\over \rho_0} \sim {\alpha\over 2} (1+z) \sim \alpha z/2
\end{equation}
It is remarkable that, for arbitrary value of $z$, all self-luminous astrophysical
objects have the same generic rule $\rho_r/\rho_0 \sim \alpha ~z$ within a factor of
few. Equally remarkable is the fact that when such objects would radiate at their respective Eddington luminosities, one would have $\rho_r/\rho_0 \sim z$, the relativistic compactness parameter. This shows the sublime role GR plays in the
structure of self-gravitating objects at all levels.
 \section{Discussion}
 If the astrophysical object is strictly non-radiating, i.e., if it is static and cold, its external spacetime is represented by the radiationless vacuum Schwarzschild
 metric. For such static cold objects, there is an upper limit of $z <2$ which is
 also called as Buchdahl limit\citep{b2,b3,b5}. However,
 it is believed that very massibe objects undergo continued collapse to
 form a Black Hole  having an Event Horizon with $z=\infty$. During collapse,
 the external spacetime is represented by radiating Vaidya metric\citep{b4} which supports
 arbitrary high value of $z$ as is required to reach the $z=\infty$ BH stage.
 During final stages of collapse, as the temperature and pressure of the body
 rise rapidly, it is likely that the body radiates at a significant fraction of its
 Eddington luminosity because $L \propto T_s^4$, where $T_s$ is the appropriate surface temperature. For the Newtonian supernova event too, we know that, peak value of $\alpha\sim 0.01$. It may however be reminded that even if one would have $\alpha\approx 1$, unlike the Newtonian case, the actually observed luminosity would
 keep on decreasing with increasing $z$ as $(1+z)^{-1}$ :
 \begin{equation}
L^\infty = {L \over (1+z)^2} = {\alpha L_{ed} \over (1+z)^2} ={4 \pi G M c m_p \alpha \over \sigma (1+z)}
\end{equation}
 Thus, when the BH would be formed ($z=\infty$), one would have 
$L^\infty =0$ even if $\alpha \sim 1$ as
 is expected. Some authors intuitively feel that the final stages
of continued gravitational collapse may be adiabatic. Note that the generic Eq.(26)
does accommodate this intuition because as $z \to \infty$, $L^\infty \to 0$.
  In any case, Eq.(25) shows that, for any finite value of $\alpha$, $\rho_r/\rho_0 \gg 1$ at appropriately high value of $z$ as one proceeds
 towards the $z=\infty$ BH stage.  Hence, during the final stages of continued GR collapse, the collapsing body is expected to become a relativistic fireball much like
 the very early stages of the universe even though a distant observer may interpret
 the collapse process as ``adiabatic'' in the sense that $L^\infty \propto (1+z)^{-1} \to 0$ as $z\to \infty$.
However, most of the GR collapse studies assume $\rho_r/rho_0 \ll 1$, and, hence, may give
incorrect conclusions. In particular, a strictly adiabatic collapse with $\rho_r =0$ would
imply $\rho_0 =0$ too.


\end{document}